\documentstyle[preprint,eqsecnum,aps,twoside]{revtex}

\newcommand{\be}{\begin{equation}}
\newcommand{\ee}{\end{equation}}
\newcommand{\bea}{\begin{eqnarray}}
\newcommand{\eea}{\end{eqnarray}}
\newcommand{\bem}{\begin{mathletters}}
\newcommand{\eem}{\end{mathletters}}
\newcommand{\sla}{\! \not \!}
\newcommand{\nn}{\nonumber}
\newcommand{\Punkt}{\hspace{3mm} .}
\newcommand{\Komma}{\hspace{3mm} ,}

\begin{document}
\draft
\preprint{SUBATECH--97--27}

\title{Relativistic Transport Theory for Systems Containing Bound States}

\author{P.~Rehberg}
\address{Institut f\"ur Theoretische Physik, Universit\"at Heidelberg, \\
         Philosophenweg 19, D--69120 Heidelberg, Germany
         \\ {\rm and} \\
         SUBATECH \\ Laboratoire de Physique Subatomique et des
         Technologies Associ\'ees \\
         UMR Universit\'e de Nantes, IN2P3/CNRS, Ecole des Mines de Nantes \\
         4 Rue Alfred Kastler, F-44070 Nantes Cedex 03, France
         \thanks{Present Address} }

\maketitle
\begin{abstract}
Using a Lagrangian which contains quarks as elementary degrees of
freedom and mesons as bound states, a transport formalism is developed,
which allows for a dynamical transition from a quark plasma to a state,
where quarks are bound into hadrons. Simultaneous transport equations
for both particle species are derived in a systematic and consistent
fashion. For the mesons a formalism is used which introduces off-shell
corrections to the off-diagonal Green functions. It is shown that these
off-shell corrections lead to the appearance of elastic quark scattering
processes in the collision integral. The interference of the processes
$q\bar q\to\pi$ and $q\bar q\to\pi\to q\bar q$ leads to a modification
of the $s$-channel amplitude of quark-antiquark scattering.
\end{abstract}


\clearpage

\section{Introduction} \label{introsec}
The search for the quark-gluon plasma (QGP) as a new state of strongly
interacting matter, where quarks and gluons become deconfined and can
move freely, has been one of the most important subjects of nuclear
and particle physics during the last decade. (For an overview see
\cite{qm95}.) While experiments to detect the QGP in heavy ion collisions
have been performed and are still performed today, the theoretical
modeling of these experiments still contains many open questions, one of
these being the description of the transition from a gas of quarks and
gluons to a gas of hadrons. While hydrodynamical models \cite{hydro} have
been successfully applied as phenomenological approaches, the derivation
of equations of motion from a microscopic Lagrangian has not yet been put
forward sufficiently in order to treat a system, that initially contains
quarks and gluons and which subsequently transforms into a system of hadrons.

Although the Schwinger--Keldysh formalism \cite{schwinke,KaBa} provides a
convenient tool for the derivation of transport equations from microscopic
models, there are still open questions to be addressed before it can be
applied to a hadronizing QGP. One imagines that the underlying Lagrangian
should, as in quantum chromodynamics (QCD), not contain hadrons as
elementary degrees of freedom, since these should appear as bound
states. On the other hand one has to consider {\em both\/} quarks and
hadrons, since the former constitute the relevant degrees in the initial,
the latter those in the final stage of the collision. Hadrons have
thus to be described as {\em effective\/} degrees of freedom and the
formalism has to allow for a {\em dynamical\/} transition between these
two regimes. This has to be accomplished in a systematic and consistent
way in order to avoid double counting.

The second requirement for the theory is that it should be possible
to derive collision terms for higher order processes. To be definite,
say e.\,g. that in lowest order one obtains an interaction between
quarks and pions via the process $q\bar q\leftrightarrow\pi$. In
higher orders also the processes $q\bar q\leftrightarrow q\bar q$
\cite{elaste} and $q\bar q\leftrightarrow\pi\pi$ \cite{hadron} give
important contributions and should be derivable within the theory. A
first sketch of this programme has been given in Ref.~\cite{hirsch}. As
it turns out, this is by far the harder task to fulfill.

The description of nonequilibrium matter in terms of quasiparticles
is in so far problematic as it is strictly valid only in the low
density limit. In Ref.~\cite{pavel} however, a formalism has been
developed which to a certain extent overcomes this problem and has been
successfully applied to semiconductor physics. The basic trick of this
approach is a decomposition of the off-diagonal Green functions of the
Schwinger--Keldysh formalism into a singular and a regular part. It
turns out that in leading order of a gradient expansion the singular
part obeys a Boltzmann equation, whereas the regular part gives
corrections to the collision terms. It has been shown already in
Ref.~\cite{pavel}, that these corrections are essential if collision
terms for higher order processes like those described above have to
be considered.

In the present paper, the approach of Ref.~\cite{pavel} is thus
applied to the dynamical formation of pions out of a quark plasma. As
a microscopic model for the interaction the Nambu--Jona-Lasinio (NJL)
model \cite{nambu,sandi} is used. Although this model fails to provide
a complete description of a QGP since it is not confining and does not
contain gluons, it has the same internal symmetries as QCD and thus gives
a good description of the low energy mesonic sector that is governed by
strong interactions. The underlying principle of this approach is
the concept of chiral symmetry breaking, which forms also the basis of
other phenomenological models such as chiral perturbation theory and
the sigma model \cite{chpt}, and which works as follows: In the limit of
vanishing current quark masses, QCD is invariant under transformations
of the form $\psi\to\exp\left(-i\theta^a\chi_a\gamma^5\right)\psi$,
where $\theta^a$ is an arbitrary vector in flavor space and $\chi_a$
the generators of the group $SU(N_f)$. This symmetry, in turn, is
dynamically broken and thus not observed in nature. As a consequence,
due to the Goldstone theorem, $N_f^2-1$ massless bosons appear, which
for $N_f=2$ are identified with the pions, and for $N_f=3$ with pions,
kaons and the $\eta$. The finite mass of these particles observed in
nature comes about due to a (small) explicit chiral symmetry breaking
by finite current quark masses. It turns out that the concept of chiral
symmetry is already sufficient to describe the low energy phenomena of
strong interactions.

In the NJL model \cite{sandi}, this is implemented as follows: one
starts with a Lagrangian containing free quarks, which interact via a
chirally symmetric contact interaction. This interaction in turn leads
to a spontaneous breakdown of chiral symmetry. As a consequence the
pions, which appear as bound states of quarks and antiquarks, become
massless. The physical picture is thus that one has pions as ground
state and constituent quarks with masses of about 300--350\,MeV. At finite
temperature the situation changes, however, since chiral symmetry is
restored at $T\approx 200$\,MeV. In this case, the quark mass drops down
to the current quark mass, while mesons become unstable resonances. Within
the present accuracy of lattice calculations, chiral symmetry is restored
at the same temperature where deconfinement happens \cite{edwin}. The
lattice data show as well, that the temperature behaviour of the meson
masses is qualitatatively the same for both lattice QCD and the NJL
model, as can be seen by comparing the data of Ref.~\cite{edwin} with
those given in Ref.~\cite{sandi}.

Due to this reasons, its simplicity and since the appearance of bound
states is well studied in equilibrium, the NJL model provides an ideal
starting point for studying the nonequilibrium behaviour of bound
states. The results obtained here may nevertheless be of larger
significance than the application to a specific model.

The derivation of transport equations for the quark sector of the NJL
model has already been done in Refs.~\cite{wojtek,ogu} and numerical
solutions have been given in Refs.~\cite{aichelin}. These works,
however, do not include the formation of bound states. An approach to
bound state formation has been attempted in Refs.~\cite{daniber,zhawi},
but without addressing the problem of computing the collision terms
for higher order processes. On the other hand, there have been several
attempts at constructing transport theories for mesonic Lagrangians that
treat mesons as point like interacting particles \cite{davis,stan}. The
present approach goes beyond these works in that it unites both the quark
and mesonic degrees of freedom in a single model, and it is required to
account for the fact that the mesonic states are bound states of quarks
below the Mott temperature \cite{gerry} and resonances above this. It is
further attempted to compute the collision terms beyond the leading order.

This paper is organized as follows: In Section~\ref{fundisec}, the
model is introduced. Since the NJL model is a strongly interacting
theory, it has to be treated by an expansion in the inverse number of
colors, $1/N_c$, rather than by an expansion in the coupling constant
\cite{expand,richard}. Thus mesons, which appear already in the lowest
order of this expansion, are introduced as effective degrees of freedom
in the beginning. In Section~\ref{quarksec}, the equations of motion
for the quark Green functions in coordinate space are obtained and
the derivation of the Boltzmann equation for the quark densities in
the standard formalism \cite{ogu}, which is sufficient for the present
purpose, is briefly reviewed. The equations of motion for the mesonic
Green functions are given in Section \ref{mesogrsec}. These are treated
further and cast into a Boltzmann form in Section~\ref{mesotrsec}.
Section~\ref{intersec} is dedicated to the explicit derivation of
collision terms. In Section~\ref{intersecA} the lowest order of the
mesonic collision term is shown to contain processes of the type
$q\bar q\leftrightarrow\pi$, which are kinematically allowed at
high temperatures. The lowest order collision integral for quarks is
computed in Section~\ref{intersecB}. Since in this collision term the
off-shell corrections to the off-diagonal meson Green functions appear,
this calculation is divided in two parts. First the contributions of
the singular (quasiparticle) part of the Green functions is shown to
yield processes of the type $q\bar q\leftrightarrow\pi$. Afterwards
the contributions of the regular part are shown to give processes of
the type $qq\leftrightarrow qq$ and $q\bar q\leftrightarrow q\bar
q$. It is essential to consider the off-shell corrections in order
to obtain these processes at all. It is, however, also shown that
the form of the off-shell corrections given in Ref.~\cite{pavel} is
insufficient in order to describe the correct $s$-channel scattering
amplitude of quark-antiquark scattering.  Conclusions are presented in
Section~\ref{conclusec}.

\section{Fundamental Equations of Motion} \label{fundisec}
The starting point for the following investigations is the NJL Lagrangian,
which in its two flavor version reads \cite{sandi}
\begin{equation}
{\cal L} = \bar \psi (i\sla\partial - m_0) \psi
    + G \left[ (\bar\psi\psi)^2 + (\bar\psi i\gamma_5\vec\tau\psi)^2 \right]
\label{LNJL}
\Punkt
\end{equation}
Here $\psi$ denotes the quark wave function, which implicitly contains
color and flavor degrees of freedom, and $\vec\tau$ are the Pauli matrices
in flavor space. The equations of motion for the quark fields obtained
from Eq.~(\ref{LNJL}) are
\begin{mathletters} \label{motion}
\begin{eqnarray}
(i\sla\partial - m_0) \psi &=&
-2G \left[ (\bar\psi\psi) + (\bar\psi i\gamma_5 \vec\tau\psi)
i\gamma_5\vec\tau \right]\psi \\
\bar\psi (-i\stackrel{\leftarrow}{\sla\partial} - m_0) &=&
-2G \bar\psi \left[ (\bar\psi\psi) + (\bar\psi i\gamma_5 \vec\tau\psi)
i\gamma_5\vec\tau \right]
\Punkt
\eea
\end{mathletters}
Sigma and pion fields can be introduced via \cite{sandi,wojtek,zhawi}
\begin{mathletters} \label{sigpi}
\begin{eqnarray}
\sigma &=& -2G\bar\psi\psi = -2G \bar\psi \Gamma_\sigma \psi \\
\pi^0 &=& -2G\bar\psi i\gamma_5 \tau_3 \psi
= -2G \bar\psi \Gamma_0\psi \\
\pi^+ &=& -2G\bar\psi i\gamma_5\frac{\tau_1 + i \tau_2}{\sqrt{2}} \psi
= -2G \bar\psi \Gamma_+\psi \\
\pi^- &=& -2G\bar\psi i\gamma_5\frac{\tau_1 - i \tau_2}{\sqrt{2}} \psi
= -2G \bar\psi \Gamma_-\psi
\Komma
\eea
\end{mathletters}
where, in order to simplify the notation in the following, also
the matrices $\Gamma_\sigma$, $\Gamma_0$ and $\Gamma_\pm$ have been
defined. Note that these definitions imply that the $\sigma$ and $\pi^0$
fields are real, whereas $\pi^+$ and $\pi^-$ form a complex field with
${\pi^+}^* = \pi^-$.

The equations of motion for the quark fields (\ref{motion}) can be
rewritten as
\begin{mathletters} \label{poetry}
\begin{eqnarray}
(i\sla\partial - m_0) \psi &=&
\left(\sigma \Gamma_\sigma + \pi^0\Gamma_0 + \pi^+\Gamma_-
+ \pi^-\Gamma_+ \right) \psi \\
\bar\psi (-i\stackrel{\leftarrow}{\sla\partial} - m_0) &=&
\bar\psi \left(\sigma \Gamma_\sigma + \pi^0\Gamma_0 + \pi^+\Gamma_-
+ \pi^-\Gamma_+ \right)
\Punkt
\eea
\end{mathletters}
In this form they can, together with the definition of the sigma and pion
fields (\ref{sigpi}), be obtained from the Lagrange density
\begin{eqnarray} \label{efflag}
{\cal L} &=& \bar\psi(i\sla\partial-m_0)\psi
         - \frac{1}{4G}\left(\sigma^2+{\pi^0}^2+2\pi^+\pi^-\right)
\\ \nn   &-& \bar\psi \left(\sigma\Gamma_\sigma
             +\pi^0\Gamma_0 +\pi^+\Gamma_-
                        +\pi^-\Gamma_+\right) \psi
\Komma
\eea
which is equivalent to the original Lagrangian (\ref{LNJL}).  A canonical
quantization of the Lagrangian (\ref{efflag}) fails however, since
(\ref{efflag}) does not contain the time derivative of the mesonic
fields and thus inhibits the definition of canonical momenta for the
meson fields.  One possible way out of this problem is to quantize
(\ref{efflag}) using the path integral approach, as is done in
Ref.~\cite{zhawi}.  Here, however, an infinitesimal kinetic energy for
the meson fields is added to the Lagrangian,
\begin{equation}
{\cal L}' = {\cal L} + \frac{1}{2}\epsilon \left[
\left(\partial_0\sigma\right)^2 +\left(\partial_0\pi^0\right)^2
+2\left(\partial_0\pi^+\right) \left(\partial_0\pi^-\right) \right]
\Komma \label{infikin}
\ee
quantization is done canonically and the limit $\epsilon\to 0$ is taken at
the end.\footnote{One could also think of adding other terms containing
spatial derivatives of the fields to the Lagangian in order to maintain
Lorentz covariance even for $\epsilon\ne 0$. However, this procedure
would not change the following calculations and for the present purpose
Eq.~(\ref{infikin}) is sufficient.} Using the generic notation $\phi_k$
for the meson fields, one then obtains the associated canonical momenta
\begin{equation}
\wp_k = \frac{\partial {\cal L}'}{\partial \left(\partial_0 \phi_k\right)}
      = \epsilon \partial_0 \phi_k^* \Komma
\ee
which allow for the introduction of canonical commutation
relations at equal time
\begin{mathletters} \label{commute}
\begin{eqnarray}
\phi_k(x) \phi_l(y) - \phi_l(y) \phi_k(x) &=& 0 \\
\wp_k(x) \wp_l(y) - \wp_l(y) \wp_k(x) &=& 0 \\
\phi_k(x) \wp_l(y) - \wp_l(y) \phi_k(x) &=&
i \delta_{kl}\delta^3(\vec x-\vec y)
\Punkt
\eea
\end{mathletters}
Using these relations, the canonical quantization of (\ref{infikin})
can proceed in a standard fashion \cite{BD}.

\subsection{Quark Fields} \label{quarksec}
The transport equations for quark fields in the NJL model are derived in
Refs.~\cite{wojtek,ogu,zhawi}. Here the main results of these works
are listed for further reference and in order to define formalism and
notation.

The Green functions for the quark fields in the Schwinger--Keldysh
formalism are defined by
\begin{mathletters} \label{gruen}
\begin{eqnarray}
S^+(x,y) &=& -i \left< {\cal T}\left[\psi(x) \bar\psi(y)\right]\right> \\
S^-(x,y) &=& -i \left< \tilde {\cal T}\left[\psi(x) \bar\psi(y)\right]\right>
\\
S^>(x,y) &=& -i \left< \psi(x) \bar\psi(y)\right> \\
S^<(x,y) &=&  i \left< \bar\psi(y)\psi(x) \right> \Komma
\eea
\end{mathletters}
where ${\cal T}$ and $\tilde{\cal T}$ are the time ordering and
anti-time ordering operators, respectively:
\begin{mathletters}
\begin{eqnarray}
{\cal T}\left[A(x)B(y)\right] &=&
\theta(x_0-y_0)A(x)B(y) \pm \theta(y_0-x_0) A(y) B(x) \\
\tilde{\cal T}\left[A(x)B(y)\right] &=&
\theta(y_0-x_0) A(x)B(y) \pm \theta(x_0-y_0) B(y)A(x) \Punkt
\eea
\end{mathletters}
Here the upper (lower) sign refers to bosonic (fermionic) operators.
In order to unify Eqs.~(\ref{gruen}), the field operators are analytically
continued to the complex time plane and the Schwinger--Keldysh contour $C$
(see Fig.~\ref{contour}) is introduced \cite{KaBa}. This contour runs
from the time $t_0$, where the initial conditions are specified, to an
arbitrary time larger than any other time considered (conveniently set
to $+\infty$) and back. The initial time $t_0$ will in the following be
shifted to $-\infty$. This can be done since one has two time scales
involved, namely the mean collision time and the expansion time. An
initial time $t_0=-\infty$ in this sense means thus a time which is large
compared to the mean collision time, but still small compared to the time
scale of the expansion. For a more formal argument for this procedure
the reader is referred to the paper of Botermans and Malfliet \cite{KaBa}.

The contour ordering operator ${\cal T}_C$ is defined as
\begin{equation}
{\cal T}_C\left[A(x) B(y)\right] = \theta_C(x_0,y_0) A(x) B(y)
\pm \theta_C(y_0,x_0) B(y) A(x) \Komma
\ee
using the contour theta function
\begin{equation}
\theta_C(x_0,y_0) = \left\{ \begin{array}{ll}
             1 & \mbox{if $x_0$ is `later'  than $y_0$ with respect to $C$} \\
             0 & \mbox{if $x_0$ is `sooner' than $y_0$ with respect to $C$}
                  \end{array} \right.
\Punkt
\ee
Using this definition, Eqs.~(\ref{gruen}) can be written as
\begin{equation}
S^C(x,y) = -i\left<{\cal T}_C\left[\psi(x) \bar\psi(y)\right]\right>
\label{pathg} \Punkt
\ee
{}From Eqs.~(\ref{efflag}) and (\ref{gruen}), one can easily
derive the equations of motion for the quark Green functions:
\begin{eqnarray} \label{KB1}
& &(i\sla\partial_x - m_0) S^C(x,y) = \delta_C(x,y)
\\ \nn & & \qquad -
i \left<{\cal T}_C\left[\left(\sigma(x)\Gamma_\sigma
+\pi^0(x)\Gamma_0+\pi^+(x)\Gamma_-+\pi^-(x)\Gamma_+
\right) \psi(x)\bar\psi(y)\right]\right>
\Komma
\eea
where the contour delta function is defined as
\begin{equation}
\delta_C(x,y) = \left\{
                \begin{array}{ll}
                \delta^4(x - y) & \mbox{if both $x_0$ and $y_0$
                                            are on the upper branch of $C$} \\
               -\delta^4(x - y) & \mbox{if both $x_0$ and $y_0$
                                            are on the lower branch of $C$} \\
                0 & \mbox{elsewhere}
                \end{array} \right. \Punkt
\ee
Defining the contour ordered self energy via
\begin{eqnarray} \label{zwei16}
& & \int_C d^4y \, \Sigma^C(x,y)S^C(y,z) =
\\ \nn & & \qquad
- i \left<{\cal T}_C\left[\left(\sigma(x)\Gamma_\sigma
+\pi^0(x)\Gamma_0+\pi^+(x)\Gamma_-+\pi^-(x)\Gamma_+
\right) \psi(x)\bar\psi(y)\right]\right>
\Komma
\eea
where the subscript $C$ at the integral sign indicates time integration
along the contour,
one can cast Eq.~(\ref{KB1}) into the Schwinger--Dyson like form
\begin{equation}
(i\sla\partial_x - m_0) S^C(x,z) = \delta_C(x,z) +
\int_C d^4y \, \Sigma^C(x,y)S^C(y,z)
\label{quadir} \Punkt
\ee
Analogously, one can derive the adjoint equation
\begin{equation}
S^C(x,z) (-i\stackrel{\leftarrow}{\sla\partial}_z - m_0) = \delta_C(x,z) +
\int_C d^4y \, S^C(x,y)\Sigma^C(y,z) \Punkt
\label{quaadj}
\ee
By defining retarded and advanced Green functions and self energies via
\bem \label{retadef} \bea
S^R(x,y) &=& S^+(x,y) - S^<(x,y) = S^>(x, y) - S^-(x, y) \\
S^A(x,y) &=& S^+(x,y) - S^>(x,y) = S^<(x, y) - S^-(x, y) \\
\Sigma^R(x,y) &=& \Sigma^+(x,y) - \Sigma^<(x,y)
               = \Sigma^>(x, y) - \Sigma^-(x, y) \\
\Sigma^A(x,y) &=& \Sigma^+(x,y) - \Sigma^>(x,y)
               = \Sigma^<(x, y) - \Sigma^-(x, y)
\eea \eem
and disentangling the contour integrations in Eqs.~(\ref{quadir}),
(\ref{quaadj}), one arrives at the equations of motion for the Green
functions
\bem \label{qkaba} \bea
\left( S_b^{-1} \otimes S^{R,A} \right)(x,y) &=& \delta^4(x-y) +
\left(\Sigma^{R,A} \otimes S^{R,A}\right) (x, y) \\
\left( S^{R,A} \otimes S_b^{-1} \right)(x,y) &=& \delta^4(x-y) +
\left( S^{R,A} \otimes \Sigma^{R,A} \right)(x,y) \\
\left( S_b^{-1} \otimes S^{>,<}\right)(x,y) &=&
\left( \Sigma^{>,<}  \otimes S^A + \Sigma^R \otimes S^{>,<} \right) (x,y) \\
\left( S^{>,<} \otimes S_b^{-1} \right)(x,y) &=&
    \left( S^{>,<} \otimes \Sigma^A + S^R \otimes \Sigma^{>,<} \right)(x,y)
\Punkt
\eea \eem
In order to abbreviate the notation, the convolution operator $\otimes$,
which is defined by
\be
\left( A \otimes B \right) (x, z) = \int d^4y \, A(x, y) B(y, z)
\label{conv}
\ee
and the inverse of the bare Green function
\be
S_b^{-1}(x, y) = (i\sla\partial_x - m_0) \delta^4(x - y)
\label{Sbare}
\ee
are introduced in Eqs.~(\ref{qkaba}).  Note that `inverse' here means
inverse with respect to the convolution (\ref{conv}):
\be
\left(F \otimes F^{-1} \right) (x, y) = \delta^4(x - y) \Punkt
\ee
The general framework for deriving transport equations from
Eqs.~(\ref{qkaba}) has been outlined in Refs.~\cite{KaBa},
an explicit calculation for the NJL model has been given in
Refs.~\cite{wojtek,ogu}. Since the aim of the present work is the
derivation of transport equations for mesons, these calculations are not
repeated here. It should however be mentioned that with the definition
of the self energy (\ref{zwei16}), $\Sigma^C(x,y)$ still contains a
Hartree part, which in the case of the NJL model leads to the spontaneous
breakdown of chiral symmetry and the appearance of constituent quarks. For
further details, the reader is referred to Refs.~\cite{sandi,wojtek}. In
Ref.~\cite{ogu}, a gradient expansion of Eqs.~(\ref{qkaba}) is performed
and subsequently the quasiparticle ansatz for $S^{>,<}$,
\begin{mathletters} \label{KBansatz} \begin{eqnarray}
S^<(x,p)
&=& \frac{i\pi}{E_q(x,\vec p)}\frac{\delta_{ff'}\delta{cc'}}{2N_cN_f}
    \left(\sla p + m_q(x)\right) \Big[ \delta\left(p_0-E_q(x,\vec p)\right)
    n_q(x, \vec p)
\\ \nonumber & & \hspace{2cm}
    - \delta\left(p_0+E_q(x,\vec p)\right) \left(2N_cN_f-n_{\bar q}
      (x, - \vec p)\right) \Big] \\
S^>(x,p)
&=& - \frac{i\pi}{E_q(x,\vec p)}\frac{\delta_{ff'}\delta{cc'}}{2N_cN_f}
    \left(\sla p + m_q(x)\right) \Big[ \delta\left(p_0-E_q(x,\vec p)\right)
    \left(2N_cN_f-n_q(x, \vec p)\right)
\nonumber \\ & & \hspace{2cm}
    - \delta\left(p_0+E_q(x,\vec p)\right) n_{\bar q} (x, - \vec p) \Big]
\Komma
\end{eqnarray} \end{mathletters}
with $E_q(x,\vec p)=\sqrt{\vec p ^2 + m_q^2(x)}$, is employed in
order to derive the Boltzmann equation
\begin{eqnarray}
& & \left[\partial_t + \vec \partial_p E_q(x, \vec p) \vec\partial_x -
\vec \partial_x E_q(x, \vec p) \vec\partial_p \right] n_q(x, \vec p) =
\label{qtrans} \\ \nonumber & & \hspace{4cm}
\int_0^\infty \frac{dp_0}{2\pi} {\rm Tr} \left[\Sigma^<(x,p)S^>(x,p)
-\Sigma^>(x,p)S^<(x,p)\right]
\end{eqnarray}
for the quark density $n_q(x, \vec p)$. The integration on the right
hand side serves to eliminate the delta functions contained implicitly
in $S^{>,<}$. The dynamical quark mass $m_q(x)$ in Eq.~(\ref{qtrans})
has in turn to be calculated from the gap equation
\begin{equation}
m_q(x) = m_0 + 2G m_q(x) \int_{|\vec p| < \Lambda}
\frac{d^3p}{(2\pi)^3} \frac{2N_cN_f-n_q(x, \vec p)-n_{\bar q} (x, \vec p)}
{E_q(x,\vec p)} \Punkt \label{gap}
\end{equation}
Together with an expansion of the nondiagonal selfenergies $\Sigma^{>,<}$,
to be discussed later, Eqs.~(\ref{qtrans}) and (\ref{gap}) provide a set
of equations for the computation of $n_q(x, \vec p)$ and
$n_{\bar q} (x, \vec p)$. A numerical evaluation of these equations has
been given in Refs.~\cite{aichelin}.

\subsection{Meson Fields} \label{mesogrsec}
In analogy to Eq.~(\ref{pathg})
the Green functions for the meson fields are defined via
\begin{mathletters} \begin{eqnarray}
\Delta^C_\sigma(x,y) &=& -i \left\{
        \left<{\cal T}_C\left[\sigma(x) \sigma(y)\right]\right>
      - \left<\sigma(x)\right> \left<\sigma(y)\right> \right\} \\
\Delta^C_0(x,y) &=& -i \left\{
        \left<{\cal T}_C\left[\pi^0(x) \pi^0(y)\right]\right>
      - \left<\pi^0(x)\right> \left<\pi^0(y)\right> \right\} \\
\Delta^C_\pm(x,y) &=& -i \left\{
        \left<{\cal T}_C\left[\pi^+(x) \pi^-(y)\right]\right>
      - \left<\pi^+(x)\right> \left<\pi^-(y)\right> \right\}
\Komma
\eea \end{mathletters}
for sigmas, neutral pions and charged pions, respectively.  Employing
the equations of motion for the fields derived from (\ref{infikin}) and
the commutation relations (\ref{commute}), one obtains the equation of
motion for the contour ordered mesonic Green functions
\begin{mathletters} \label{vomit} \begin{eqnarray}
&-&\left(\epsilon \partial_{x_0}^2 + \frac{1}{2G}\right) \Delta^C_\sigma(x,y) =
\delta_C(x,y) \\ \nn & & \hspace{2cm} - i \left\{\left<{\cal T}_C\left[
\bar\psi(x)\Gamma_\sigma\psi(x) \sigma(y)\right]\right>
- \left<\bar\psi(x)\Gamma_\sigma\psi(x)\right>\left<\sigma(y)\right> \right\}
\\
&-&\left(\epsilon \partial_{x_0}^2 + \frac{1}{2G}\right) \Delta^C_0(x,y) =
\delta_C(x,y) \\ \nn & & \hspace{2cm} - i \left\{\left<{\cal T}_C\left[
\bar\psi(x)\Gamma_0\psi(x) \pi^0(y)\right]\right>
- \left<\bar\psi(x)\Gamma_0\psi(x)\right>\left<\pi^0(y)\right> \right\} \\
&-&\left(\epsilon \partial_{x_0}^2 + \frac{1}{2G}\right) \Delta^C_\pm(x,y) =
\delta_C(x,y) \\ \nn & & \hspace{2cm} - i \left\{\left<{\cal T}_C\left[
\bar\psi(x)\Gamma_+\psi(x) \pi^-(y)\right]\right>
- \left<\bar\psi(x)\Gamma_+\psi(x)\right>\left<\pi^-(y)\right> \right\}
\Punkt
\eea \end{mathletters}
Defining the contour ordered polarization function $\Pi^C_\sigma(x,y)$ for the
sigma field via
\begin{eqnarray}
& &\int_C d^4y \, \Pi^C_\sigma(x,y)\Delta^C_\sigma(y,z) =
\\ \nn & & \hspace{2cm}
-i \left\{\left<{\cal T}_C\left[
\bar\psi(x)\Gamma_\sigma\psi(x) \sigma(z)\right]\right>
- \left<\bar\psi(x)\Gamma_\sigma\psi(x)\right>\left<\sigma(z)\right> \right\}
\Komma
\eea
and analogously the polarization functions $\Pi^C_0(x,y)$ for neutral pions
and $\Pi^C_\pm(x,y)$ for charged pions,
Eqs.~(\ref{vomit}) can be rewitten as
\be
-\left(\epsilon \partial_{x_0}^2 + \frac{1}{2G}\right) \Delta^C_k(x,z) =
\delta_C(x,z)
+ \int_C d^4y \, \Pi^C_k(x,y)\Delta^C_k(y,z)
\label{mesdir}
\ee
for $k = \sigma$, $0$, $\pm$.
The adjoint equation is
\be
-\left(\epsilon \partial_{z_0}^2 + \frac{1}{2G}\right) \Delta^C_k(x,z) =
\delta_C(x,z)
+ \int_C d^4y \, \Delta^C_k(x,y)\Pi^C_k(y,z)
\Punkt
\label{mesadj}
\ee
In the limit $\epsilon\to 0$, the time derivative on the left hand side
of Eqs.~(\ref{mesdir}) and (\ref{mesadj}) drops out and one reobtains
the equations of motion of Ref.~\cite{zhawi}.

Disentangling the contour integration and introducing retarded and
advanced quantities as in Eq.~(\ref{retadef}) leads to
\bem \label{KBmes} \bea
\left( \Delta_b^{-1} \otimes \Delta_k^{R,A} \right) (x, y) &=& \delta^4(x-y) +
\left( \Pi_k^{R,A} \otimes \Delta_k^{R,A} \right) (x,y) \label{retm1}
\\
\left( \Delta_k^{R,A} \otimes \Delta_b^{-1} \right) (x, y) &=& \delta^4(x-y) +
\left( \Delta_k^{R,A} \otimes \Pi_k^{R,A} \right) (x,y) \label{retm2}
\\
\left( \Delta_b^{-1} \otimes \Delta_k^{>,<} \right) (x, y) &=&
\left( \Pi_k^{>,<} \otimes \Delta_k^A
+ \Pi_k^R \otimes \Delta_k^{>,<}\right) (x,y) \label{ndm1}
\\
\left( \Delta_k^{>,<} \otimes \Delta_b^{-1} \right) (x, y) &=&
\left( \Delta_k^{>,<} \otimes \Pi_k^A
+ \Delta_k^R \otimes \Pi_k^{>,<}\right) (x,y) \label{ndm2} \Komma
\eea \eem
where the inverse of the bare propagator is given by
\be
\Delta_b^{-1}(x,y) = -\frac{1}{2G} \delta(x-y) \Punkt
\label{Dbare}
\ee
Equations (\ref{KBmes}) are the mesonic analogue of Eqs.~(\ref{qkaba})
and will form the basis of the following investigations. A major
difference between Eqs.~(\ref{KBmes}) and the corresponding equations
for particles, which occur as elementary degrees of freedom in the
microscopic interaction model, is that the inverse bare propagator
defined in Eq.~(\ref{Dbare}) does no longer contain derivatives, as does
e.\,g. the inverse bare quark propagator given in Eq.~(\ref{Sbare}).
It will be seen in the following, however, that this does not inhibit
the description of mesons within the framework of transport theory.

\section{Transport Equations for Mesons} \label{mesotrsec}
The derivation of transport equations from Eqs.~(\ref{KBmes}) can in
principle be done in the standard fashion outlined in
Refs.~\cite{KaBa,ogu}. The drawback of this approach is, however,
that the quasiparticle ansatz, which for bosons reads
\bem \label{kbmeson} \bea
\Delta_k^<(x,p) &=& - \frac{i\pi}{E_k(x, \vec p)}\Big[n_k(x, \vec p)
\delta(p_0-E_k(x, \vec p))
\\ \nonumber & & \hspace{2cm}
+ (1+n_{\bar k}(x, - \vec p))\delta(p_0+E_k(x, \vec p)) \Big]
\\
\Delta_k^>(x,p) &=& - \frac{i\pi}{E_k(x, \vec p)}\Big[(1+n_k(x, \vec p))
\delta(p_0-E_k(x, \vec p))
\nonumber \\ & & \hspace{2cm}
+ n_{\bar k}(x, - \vec p)\delta(p_0+E_k(x, \vec p)) \Big] \Komma
\eea \eem
does not solve the equations of motion (\ref{ndm1}), (\ref{ndm2}) unless
one neglects off-shell terms \cite{pavel}. Since the problem treated
here differs from the standard formalism in considering the propagation
of non-elementary particles, it is not clear a priori, whether this
procedure is justified. It is thus desirable to use a formalism which
consistently eliminates off-shell terms within the framework of the
gradient expansion, as has been developed in Ref.~\cite{pavel}. It
has been shown there, that the unwanted terms can be eliminated by
considering off-shell corrections to the ansatz (\ref{kbmeson}).  Thus,
in the following the formalism developed in \cite{pavel} will be applied
to (\ref{KBmes}).

\subsection{Mesonic Spectral Functions}
In order to transform Eqs.~(\ref{KBmes}) to momentum space, a Wigner
transformation is performed, which for a function $F(x,y)$ of two
coordinate arguments is defined by
\be
F(x, p) = \int d^4u \, e^{ipu} F(x + u/2, x - u/2) \Punkt
\ee
The Wigner transform of a convolution of two functions can be evaluated
employing a gradient expansion \cite{KaBa}. Keeping only terms up to
first order gradients, one obtains
\bea \label{gradex}
& & \int d^4u \, e^{ipu} \left( F \otimes G \right) (x + u/2, x - u/2)
\\ \nn & & \hspace{4cm}
= F(x,p) G(x,p) + \frac{i}{2}\left\{F(x,p); G(x,p) \right\} \Komma
\eea
introducing the Poisson bracket of two functions in momentum space,
\be
\left\{F(x,p); G(x,p) \right\} = \partial_p F(x,p) \partial_x G(x,p)
- \partial_x F(x,p) \partial_p G(x,p) \Punkt
\ee
In the following, also the Wigner transform of triple and quadruple
convolutions will be needed, which can be easily inferred from
Eq.~(\ref{gradex}).

By transforming Eqs.~(\ref{retm1}) and (\ref{retm2}) and taking
the sum and the difference of the transformed equations, one obtains
\bem \label{Dr} \begin{eqnarray}
-\frac{1}{2G} \Delta_k^{R,A}(x,p) &=& 1 + \Pi_k^{R,A}(x,p) \Delta_k^{R,A}(x,p)
\\
0 &=& \left\{ \Pi_k^{R,A} (x,p); \Delta_k^{R,A}(x,p) \right\}
\Punkt
\eea \eem
Equations (\ref{Dr}) are solved by
\be
\Delta_k^{R,A}(x,p) = - \frac{2G}{1+2G\Pi_k^{R,A}(x,p)} \label{retsol}
\Punkt
\ee
The mesonic quasiparticle energy and the width can be identified as the
location of the poles of the retarded Green function (\ref{retsol}),
i.\,e. as the solution of the dispersion relation
\be
1+2G\Pi_k^R(x,p)=0 \Punkt \label{dispers}
\ee
In the vicinity of a pole, the denominator of Eq.~(\ref{retsol})
can be Taylor-expanded with respect to $p_0$ as
\be
1+2G\Pi_k^R(x,p) \approx - \frac{2G}{g_k^2(x, \vec p) - i a_k(x, \vec p)}
2E_k(x,\vec p)
\left( p_0 - E_k(x,\vec p) + \frac{i}{2} \Gamma_k(x,\vec p) \right)
\Komma \label{quasip}
\ee
defining the quasiparticle energy $E_k(x,\vec p)$, the quasiparticle width
$\Gamma_k(x,\vec p)$ and the effective meson-quark coupling $g_k(x, \vec
p)$ for meson species $k$. With Eq.~(\ref{quasip}), $\Delta_k^R(x, p)$
becomes in the vicinity of the pole
\be
\Delta_k^R(x,p) \approx
\frac{g_k^2(x, \vec p) - i a_k(x, \vec p)}{2 E_k(x,\vec p)
\left( p_0 - E_k(x,\vec p) + \frac{i}{2} \Gamma_k(x,\vec p) \right)}
\label{pola}
\ee
and the spectral function becomes
\bea
\rho_k(x,p) &=& i \left(\Delta_k^R(x,p) - \Delta_k^A(x,p) \right)
\label{specfun} \\ \nonumber
&\approx& \frac{1}{E_k(x,\vec p)}\frac{\frac{1}{2}\Gamma_k(x,\vec p)
g_k^2(x, \vec p) + (p_0-E_k(x,\vec p)) a_k(x, \vec p)}
{(p_0-E_k(x,\vec p))^2 + \frac{1}{4}\Gamma_k^2(x,\vec p)} \Punkt
\eea
Note that $\rho_k$ is a real function, since in momentum space
$\Delta_k^A$ is the complex conjugate of $\Delta_k^R$, as can
be inferred from its definition \cite{KaBa}.

It is easy to see that the imaginary part of the pole residue, $a_k$,
is of the same order as the width $\Gamma_k$. Thus, in the limit
$\Gamma_k\to 0$ one obtains $\rho_k \sim \delta(p_0 - E_k)$, i.\,e. the
mesons are strictly on shell. The expansions (\ref{pola}) and
(\ref{specfun}) contain only a particle pole at positive $p_0$,
whereas one should expect also an antiparticle pole at negative $p_0$
from symmetry reasons \cite{davis}. Since these poles are separated by
a finite gap, however, the form given above will be sufficient.

For illustration, the spectral function for the pion is shown in
Figs.~\ref{spekt1} and \ref{spekt2} for thermal equilibrium and $\vec p=0$
at $T=0$ and $T=300$~MeV, respectively, in the random phase approximation
\cite{sandi}. In this case, the spectral function is a function of
$p_0$ only. One can distinguish two cases: at low temperatures as in
Fig.~\ref{spekt1}, Eq.~(\ref{dispers}) has solutions for real $|p_0| <
2m_q$. In this case the spectral function has the form of a delta function
in the vicinity of the solution of Eq.~(\ref{dispers}), i.\,e. contains
contributions of bound states.  Furthermore, one observes the appearance
of a continuum at $|p_0|>2m_q$, which stems from quark--antiquark
scattering states.

At high temperatures, Eq.~(\ref{dispers}) has no longer
solutions for real, but only for complex $p_0$ with $|\Re (p_0)| >
2m_q$. Physically this means that the pion becomes unstable at high
temperatures. Details about this so-called Mott transition can be
found in Ref.~\cite{gerry}. The spectral function at a temperature of
$T=300$~MeV is shown in Fig.~\ref{spekt2}. One notices, that the pion does
not manifest itself by a delta peak any longer, but rather by a resonance
peak. The dashed line in Fig.~\ref{spekt2} shows the pole approximation
according to the second line of Eq.~(\ref{specfun}), which can be seen
to give a good approximation to the full spectral function even at this
rather high temperature. Note that it is important to include these
resonant states in the theory since they give a contribution to the
effective number of degrees of freedom in the thermodynamical quantities,
as has been shown in Ref.~\cite{su2thermo}.

\subsection{Mesonic Densities} \label{mdersec}
The determination of $\Delta_k^{R,A}$ allows one to extract information
about the properties of mesons. These functions do not, however,
give direct information about the particle densities. These must be
extracted from the off-diagonal Green functions $\Delta_k^{>,<}$. As
was stated above, the approach developed in Ref.~\cite{pavel} will be
applied. This formalism starts by returning to the equations of motion
in coordinate space and observing that Eqs.~(\ref{ndm1}), (\ref{ndm2})
are equivalent to \cite{pavel}
\be \label{deltasol}
\Delta_k^{>,<} = \Delta_k^R  \otimes \Pi_k^{>,<} \otimes \Delta_k^A \Komma
\ee
where the coordinate arguments have been dropped for simplicity.
Analogously, $\rho_k$ can be rewritten as
\be \label{rhosol}
\rho_k = i \left(\Delta_k^R - \Delta_k^A\right)
             =  i \left(\Delta_k^> - \Delta_k^<\right)
            = \Delta_k^R \otimes \gamma_k \otimes \Delta_k^A \Punkt
\ee
In Eq.~(\ref{rhosol}), the auxiliary quantities
\bem \label{pidecomp} \bea
\Pi_k &=& \frac{1}{2} \left(\Pi_k^R + \Pi_k^A \right) \\
\gamma_k &=& i \left(\Pi_k^R - \Pi_k^A \right)
          =  i \left(\Pi_k^> - \Pi_k^< \right)
\eea \eem
have been introduced. Since in momentum space the retarded and
advanced quantities are complex conjugate to each other \cite{KaBa},
Eq.~(\ref{pidecomp}) is for this case commensurate with a decomposition
of $\Pi_k^{R,A}$ into its real and imaginary part.

Starting from the identities (\ref{deltasol}) and (\ref{rhosol}), it is
possible to decompose $\Delta_k^{>,<}$ and $\rho_k$ into a singular and
a regular part
\bem \label{drei12} \bea
\Delta_k^{>,<} &=& \Delta_{k,s}^{>,<} + \Delta_{k,r}^{>,<} \\
\rho_k &=& \rho_{k,s} + \rho_{k,r}
\Komma
\eea \eem
which in turn are defined by
\bem \label{drei13} \bea
\Delta_{k,s}^{>,<} &=& \frac{i}{2} \left(
\Delta_k^R  \otimes \Pi_k^{>,<} \otimes \rho_k
- \rho_k \otimes \Pi_k^{>,<} \otimes \Delta_k^A \right) \label{singdef}
\\
\Delta_{k,r}^{>,<} &=& \frac{1}{2} \left(
\Delta_k^R \otimes \Pi_k^{>,<} \otimes \Delta_k^R
+ \Delta_k^A \otimes \Pi_k^{>,<} \otimes \Delta_k^A \right) \\
\rho_{k,s} &=& \frac{i}{2} \left[
\Delta_k^R \otimes \gamma_k \otimes \rho_k
- \rho_k \otimes \gamma_k \otimes \Delta_k^A
\right] \label{rhosing}
\\
\rho_{k,r} &=& \frac{1}{2} \left[
\Delta_k^R \otimes \gamma_k \otimes \Delta_k^R
+ \Delta_k^A \otimes \gamma_k \otimes \Delta_k^A
\right] \Punkt
\eea \eem

The physical implications of the decompositions (\ref{drei12})
and (\ref{drei13}) have been discussed in Ref.~\cite{pavel}. The
decomposition of $\Delta_k^{>,<}$ and $\rho_k$ into a singular and
regular part corresponds to a separation of long time and short time
phenomena. It is thus possible to describe the singular part (which
corresonds to the long time phenomena) in terms of quasiparticles.
This picture is further confirmed by observing that by Wigner transforming
Eq.~(\ref{rhosing}) and computing the gradients using Eq.~(\ref{retsol}),
one obtains $\rho_{k,s} = \frac{1}{2}\gamma_k\rho_k^2$ in momentum space,
i.\,e. for a given width the singular part of the spectral function
$\rho_{k,s}$ forms a much narrower peak around the quasiparticle pole
than the spectral function itself.

If a description of mesons in terms of quasiparticles is desired, one
thus needs an equation of motion for the singular part of $\Delta_k^{>,<}$
{\em only\/}. Following Ref.~\cite{pavel}, this is achieved by computing
\be
-i \left[ \left(\Delta_k^R\right)^{-1} \otimes \Delta_{k,s}^<
- \Delta_{k,s}^< \otimes  \left(\Delta_k^A\right)^{-1} \right]
\label{2ways}
\ee
in two ways: Inserting
$\left(\Delta_k^{R,A}\right)^{-1} = \Delta_b^{-1} - \Pi_k^{R,A}$
 from Eqs.~(\ref{retm1}), (\ref{retm2}) leads to
\bea
& &-i \left[ \left(\Delta_k^R\right)^{-1} \otimes \Delta_{k,s}^<
- \Delta_{k,s}^< \otimes  \left(\Delta_k^A\right)^{-1} \right]
= \frac{1}{2} \left( \gamma_k \otimes \Delta_{k,s}^<
+ \Delta_{k,s}^< \otimes \gamma_k \right)
\label{way1} \\ \nonumber & & \hspace{3cm}
 - i \left[ \left( \Delta_b^{-1} - \Pi_k \right) \otimes \Delta_{k,s}^<
 - \Delta_{k,s}^< \otimes \left( \Delta_b^{-1} - \Pi_k \right) \right]
\Punkt
\eea
On the other hand, inserting Eq.~(\ref{singdef}) and
Eq.~(\ref{rhosol}) into (\ref{2ways}) yields
\bea
& &-i \left[ \left(\Delta_k^R\right)^{-1} \otimes \Delta_{k,s}^<
- \Delta_{k,s}^< \otimes  \left(\Delta_k^A\right)^{-1} \right]
= \frac{1}{2} \left( \Pi_k^< \otimes \rho_k + \rho_k \otimes \Pi_k^< \right)
\label{way2} \\ \nonumber & & \hspace{3cm}
- \frac{1}{2} \left(
\gamma_k \otimes \Delta_k^A \otimes \Pi_k^< \otimes \Delta_k^A
+ \Delta_k^R \otimes \Pi_k^< \otimes \Delta_k^R \otimes \gamma_k \right)
\Punkt
\eea
Equating (\ref{way1}) and (\ref{way2}) and performing a Wigner
transformation gives
\be
\left\{ \Delta_b^{-1} - \Pi_k ; \Delta_{k,s}^< \right\}
- \frac{1}{4} \left\{ \gamma_k ;
\rho_k \Pi_k^< \left( \Delta_k^R + \Delta_k^A \right) \right\}
= i \left( \Pi_k^< \Delta_{k,s}^> - \Pi_k^> \Delta_{k,s}^< \right) \Komma
\label{precursor}
\ee
where now all functions depend on $x$ and $p$. In deriving
Eq.~(\ref{precursor}), it has been used that
\bea
\rho_{k,s}(x,p) &=& \rho_k(x,p) - \rho_{k,r}(x,p) \\ \nonumber
                &=& \rho_k(x,p) - \frac{1}{2}\gamma_k(x,p)
                    \left( \Delta_k^R(x,p)^2 + \Delta_k^A(x,p)^2 \right)
\Punkt
\eea
Equation (\ref{precursor}) already contains the ingredients of the
Boltzmann equation. A term similar to the first one on the left hand
side of Eq.~(\ref{precursor}) appears also in the standard formalism,
where it gives rise to a mean field term. The second term on the left
hand side is particular to the present approach and has to be treated
further. The right hand side has already the form of a collision term.

In accordance with the assumption that $\Delta_{k,s}^{>,<}$ can be
described in terms of quasiparticles, these functions are factorized as
\bem \bea
\Delta_{k,s}^<(x, p) &=& -i \rho_{k,s}(x, p) f_k(x, p) \\
\Delta_{k,s}^>(x, p) &=& -i \rho_{k,s}(x, p) \left[1+f_k(x, p)\right] \Komma
\eea \eem
where $f_k$ is assumed to be a {\em smooth\/} function of $p_0$. It is
necessary to use the same function $f_k$ for both $\Delta_{k,s}^<$ and
$\Delta_{k,s}^>$ in order to maintain the identity $i\left(\Delta_{k,s}^>
- \Delta_{k,s}^<\right) = \rho_{k,s}$. The left hand side of
Eq.~(\ref{precursor}) becomes
\bea
& & \left\{ \Delta_b^{-1} - \Pi_k ; \Delta_{k,s}^< \right\}
- \frac{1}{4} \left\{ \gamma_k ;
\rho_k \Pi_k^< \left( \Delta_k^R + \Delta_k^A \right) \right\}
= -i \rho_{k,s} \left\{ \Delta_b^{-1} - \Pi_k ; f_k \right\}
\\ \nonumber & & \hspace{3cm}
- i f_k \left\{ \Delta_b^{-1} - \Pi_k ; \rho_{k,s} \right\}
- \frac{1}{4} \left\{ \gamma_k ;
\rho_k \Pi_k^< \left( \Delta_k^R + \Delta_k^A \right) \right\}
\label{rhs1} \Punkt
\eea
The second term on the right hand side can be transformed to \cite{pavel}
\be \label{drei23}
\left\{ \Delta_b^{-1} - \Pi_k ; \rho_{k,s} \right\} = \left\{ \gamma_k ;
\frac{1}{4} \gamma_k \rho_k \left( \Delta_k^R + \Delta_k^A \right) \right\}
\Punkt
\ee
With this, one can rewrite the right hand side of Eq.~(\ref{rhs1}) as
\bea
& & -i \rho_{k,s} \left\{ \Delta_b^{-1} - \Pi_k ; f_k \right\}
- \left\{ \gamma_k ; \frac{1}{4} \rho_k \left( \Delta_k^R + \Delta_k^A \right)
\left[ i \gamma_k f_k + \Pi_k^< \right] \right\}
\\ \nonumber & & \hspace{4cm}
+ \frac{i}{4} \gamma_k \rho_k \left( \Delta_k^R + \Delta_k^A \right)
\left\{ \gamma_k ; f_k \right\} \Punkt
\eea
The second term here is effectively of second order in the gradients and
can thus be neglected. The third term can be rewritten to be \cite{pavel}
\be
\frac{1}{4} \gamma_k \rho_k \left( \Delta_k^R + \Delta_k^A \right)
\left\{ \gamma_k ; f_k \right\} = \rho_{k,s} \frac{\Delta_b^{-1} - \Pi_k}
{\gamma_k} \left\{ \gamma_k ; f_k \right\} \Komma
\ee
so that Eq.~(\ref{precursor}) becomes
\be
-i \rho_{s,k} \left[ \left\{ \Delta_b^{-1} - \Pi_k ; f_k \right\}
- \frac{\Delta_b^{-1} - \Pi_k}
{\gamma_k} \left\{ \gamma_k ; f_k \right\} \right]
= \rho_{s,k} \left[  \Pi_k^< (1 + f_k) - \Pi_k^> f_k \right] \Punkt
\label{mesoboltz}
\ee
Since $\rho_{k,s}$ is strongly peaked near the quasiparticle energy,
it is possible to drop the common factor $\rho_{k,s}$ and to consider
the rest of Eq.~(\ref{mesoboltz}) on the mass shell $p_0=E_k(x, \vec
p)$. This corresponds to approximating the singular part of the spectral
function by a delta function:
\be
\rho_{s,k}(x, p) \approx
\frac{\pi g_k^2}{E_k(x, \vec p)} \left[ \delta(p_0 - E_k(x, \vec p))
- \delta(p_0 + E_k(x, \vec p)) \right] \Punkt
\ee
This in turn allows the introduction of the quasiparticle ansatz
(\ref{kbmeson}) as an ansatz {\em for the singular part of
$\Delta_k^{>,<}$ only\/}:
\bem \label{slmeson} \bea
\Delta_{k,s}^<(x,p) &\approx& - g_k^2(x, \vec p)
\frac{i\pi}{E_k(x, \vec p)} \Big[ n_k(x, \vec p) \delta(p_0-E_k(x, \vec p))
\\ \nonumber & & \hspace{4cm}
+ (1+n_{\bar k}(x, - \vec p)) \delta(p_0 + E_k(x, \vec p)) \Big]
\\
\Delta_{k,s}^>(x,p) &\approx& - g_k^2(x, \vec p)
\frac{i\pi}{E_k(x, \vec p)}\Big[(1+n_k(x, \vec p)) \delta(p_0-E_k(x, \vec p))
\\ \nonumber & & \hspace{4cm}
+ n_{\bar k}(x, - \vec p) \delta(p_0 + E_k(x, \vec p)) \Big]
\Punkt
\eea \eem
These expressions differ from Eqs.~(\ref{kbmeson}) by the factor
$g_k^2$, which stems from the pole residue of the retarded Green
functions, cf. Eq.~(\ref{pola}). Similar factors appear also in the
standard formalism, where they account for wave function renormalization
\cite{pavel}. Equation (\ref{KBansatz}) contains no such factor since the
quark self energy in the Hartree approximation is momentum independent.

{}From Eqs.~(\ref{slmeson}), one obtains $f_k(x,p)=n_k(x, \vec p)$
at positive $p_0$. Inserting this into the left hand side of
Eq.~(\ref{mesoboltz}) and using the expansion (\ref{quasip}) yields
\bea
\left\{ \Delta_b^{-1} - \Pi ; f_k \right\} &=&
\left\{ \frac{2E_k}{g_k^4+a_k^2} \left[ g_k^2 (p_0-E_k) - \frac{1}{2}
a_k \Gamma_k \right] ; n_k \right\}
\\ \nonumber &=&
\frac{2E_k}{g_k^2} \left[ \partial_t + \vec \partial_p E_k \vec \partial_x
- \vec \partial_x E_k \vec \partial_p \right] n_k \Komma
\eea
where the terms containing $a_k$ have been neglected, since they are
of second order in the width and thus beyond the validity of the
approach.

Within the same accuracy, the second term on the left hand side of
Eq.~(\ref{mesoboltz}) vanishes and one obtains
\be \label{drei30}
\left[ \partial_t
+ \vec \partial_p E_k \vec \partial_x - \vec \partial_x E_k
\vec \partial_p \right] n_k = i \frac{g_k^2}{2E_k}
\left[ \Pi_k^< (1 + n_k) - \Pi_k^> n_k \right] \Komma
\ee
i.\,e. the Boltzmann equation for mesons.

Equation (\ref{drei30}) is one of the major results of this paper. It
shows, that transport theory is applicable to the formation of bound
states and that the mean field term, i.\,e. the left hand side of
Eq.~(\ref{drei30}), has precisely the same form as it has for elementary
particles. This means that in a situation, where no free quarks are
present, mesons can propagate freely and can also carry kinetic energy. At
a first glance, this might be surprising, since the left hand sides
of Eqs.~(\ref{KBmes}) do not contain any derivatives, which usually
are associated with the kinetic energy of the particles \cite{KaBa}.
This is nevertheless only true for models, where the interaction appears
as a small perturbation and the drift term has to be present already for
the noninteracting case. Bound states, on the other hand, are {\em
nonperturbative\/} phenomena, for which the presence of interactions
is essential. The interaction part of the retarded and advanced Green
functions does thus gain a greater importance than the bare part, giving
rise to poles which are not present in the noninteracting case. These
poles, in turn, define the kinetic energy which later appears in the
drift term of the transport equations.

\section{Interactions Between Quarks and Mesons} \label{intersec}
In the last two sections, the Boltzmann equations for quark and meson
degrees of freedom have been derived and an explicit form of the mean
field (Vlasov) part has been given. It remains to evaluate the self
energies in order to express the collision integrals also in terms of
the particle densities.  In this section, the lowest order terms for the
off diagonal polarization $\Pi_k^{>,<}$ and self energies $\Sigma^{>,<}$
are examined. It is shown that these diagrams describe (i) the decay of
mesons into quark-antiquark pairs and the corresponding recombination,
$M \leftrightarrow q\bar q$, and (ii) the elastic scattering of quarks
and antiquarks via meson exchange, $qq \leftrightarrow qq$ and $q\bar
q\leftrightarrow q\bar q$. It turns out that the latter processes stem
from the off-shell part of $\Delta_k^{>,<}$ and that the interference
of the processes $q\bar q\to M$ and $q\bar q\to q\bar q$ leads to a
modified $s$-channel amplitude. The inclusion of off-shell corrections
to $\Delta_k^{>,<}$ is thus essential for a description of collisions
beyond the leading order.

Throughout this section it will be assumed that the quasiparticle energies
can be parametrized as $E(x,\vec p)=\sqrt{\vec p^2 + m^2(x)}$.

\subsection{Collision Term for Mesons} \label{intersecA}
The lowest order term for the polarization in a $1/N_c$ expansion is
shown in Fig.~\ref{polafig}.  For definiteness, the neutral pion is
considered in the following, i.\,e.  $\Gamma_k = i\gamma_5\tau_3$. The
off-diagonal polarization $\Pi_0^<$ is given by
\bea
-i\Pi_0^<(x, y) &=&
- {\rm Tr} \left[ i\Gamma_0 \, iS^C(x, y) \, i\Gamma_0 \, iS^C(y, x) \right]^<
\\ \nonumber &=&
- {\rm Tr} \left[ i\Gamma_0 \, iS^<(x, y) \, i\Gamma_0 \, iS^>(y, x) \right]
\Punkt
\eea
After Wigner transformation this becomes
\bea
\label{vollpolar} & & -i\Pi_0^<(x,p)
\\ \nonumber & & \hspace{1cm}
= - \int \frac{d^4p_1}{(2\pi)^4} \frac{d^4p_2}{(2\pi)^4}
(2\pi)^4 \delta^4(p - p_1 - p_2)
{\rm Tr} \left[ i\Gamma_0 \, iS^<(x, p_1) \, i\Gamma_0 \, iS^>(x, -p_2) \right]
\Punkt
\eea
Inserting the quasiparticle form (\ref{KBansatz}) into (\ref{vollpolar})
and multiplying out the contributions of quarks and antiquarks leads
to the appearance of four terms, which correspond to the processes
$q\to q\pi^0$, $\bar q\to\bar q\pi^0$, $\emptyset\to q\bar q\pi^0$
(i.\,e. a quark, an antiquark and a pion are emitted spontaneously from
the vacuum) and $q \bar q\to\pi^0$. This calculation proceeds along the
same lines as given in Ref.~\cite{ogu}.  Since all momenta have to be
taken on shell, the contributions of the first three processes vanish
due to kinematical constraints. The last process is allowed if $m_\pi >
2m_q$, i.\,e. for sufficiently high temperatures. Its contribution reads,
dropping the coordinate argument $x$,
\bea
& &-i\Pi_0^<(p) = \int \frac{d^4p_1}{(2\pi)^4} \frac{d^4p_2}{(2\pi)^4}
(2\pi)^4\delta^4(p-p_1-p_2) {\rm tr} \left[ \gamma_5 (\sla p_1 + m_q)
\gamma_5 (-\sla p_2 + m_q) \right]
\\ \nonumber & & \hspace{1cm} \times
\frac{(-i\pi)^2}{N_c N_f \, 2E_q(\vec p_1) \, 2E_q(\vec p_2)}
\delta(p_{10}-E_q(\vec p_1))\delta(p_{20}-E_q(\vec p_2))
n_q(\vec p_2) n_{\bar q}(\vec p_2) \Punkt
\eea
Using the on-shell constraints for both quarks and pions, this can be
simplified to
\bea
& & -i\Pi_0^<(p) = \label{terror}
- \frac{1}{g_\pi^2} \int \frac{d^3p_1}{(2\pi)^3 2E_q(\vec p_1)}
\frac{d^3p_2}{(2\pi)^3 2E_q(\vec p_2)} (2\pi)^4 \delta^4(p-p_1-p_2)
\\ \nonumber & & \hspace{3cm} \times
\overline{\left| {\cal M} \right|^2} n_q(\vec p_1) n_{\bar q}(\vec p_2)
\Komma
\eea
where
\be
\label{tramp}
\overline{\left| {\cal M} \right|^2} = \frac{m_\pi^2g_\pi^2}{2N_cN_f}
\ee
is the transition amplitude for the process $q\bar q\to\pi$. Note that
the quantities on the right hand side itself depend on the state of the
system. By a similar calculation as for Eq.~(\ref{terror}), $\Pi_0^>$
can be evaluated to be
\bea
& & -i\Pi_0^>(p) = \label{horror}
- \frac{1}{g_\pi^2} \int \frac{d^3p_1}{(2\pi)^3 2E_q(\vec p_1)}
\frac{d^3p_2}{(2\pi)^3 2E_q(\vec p_2)} (2\pi)^4 \delta^4(p-p_1-p_2)
\\ \nonumber & & \hspace{3cm} \times
\overline{\left| {\cal M} \right|^2} (2N_cN_f - n_q(\vec p_1)) (2N_cN_f -
n_{\bar q}(\vec p_2))
\eea
with $\overline{\left| {\cal M} \right|^2}$ given by Eq.~(\ref{tramp})
\footnote{Note that the transition amplitudes for $q\bar q\to \pi^0$
and $\pi^0\to q\bar q$ differ due to the different averaging factors
for the incoming particles. If the transition amplitude for the process
$\pi^0\to q\bar q$ would be used in Eq.~(\ref{horror}), the blocking
factors would have to be replaced with $1-n_q(\vec p)/(2N_cN_f)$.}.
Inserting (\ref{terror}) and (\ref{horror}) into (\ref{drei30}) yields
the full Boltzmann equation:
\bea
& &\left[ \partial_t + \vec \partial_p E_\pi \vec \partial_x
- \vec \partial_x E_\pi \vec \partial_p \right] n_\pi(\vec p)
\\ \nonumber & & \hspace{7mm}
= \frac{1}{2E_\pi}
 \int \frac{d^3p_1}{(2\pi)^3 2E_q(\vec p_1)}
\frac{d^3p_2}{(2\pi)^3 2E_q(\vec p_2)} (2\pi)^4 \delta^4(p-p_1-p_2)
\overline{\left| {\cal M} \right|^2}
\\ \nonumber & & \hspace{17mm} \times
\left\{n_q(\vec p_1) n_{\bar q}(\vec p_2) [1 + n_\pi(\vec p)] -
n_\pi(\vec p)[2N_cN_f - n_q(\vec p_1)][2N_cN_f - n_{\bar q}(\vec p_2)] \right\}
\Punkt
\eea
This equation has exactly the form expected for a quark--meson plasma
interacting via the decay of mesons into quarks and the recombination
of quarks into mesons.

\subsection{Collision Term for Quarks} \label{intersecB}
The two lowest order diagrams for the quark self energy in a $1/N_c$
expansion are shown in Fig.~\ref{sigmafig}. The diagram shown in
Fig.~\ref{sigmafig}a corresponds to the Hartree self energy and does
not contribute to the collision term \cite{ogu}, so that the lowest
order contribution to the collision term stems from the diagram shown
in Fig.~\ref{sigmafig}b. The contribution of this diagram is given by
\bea
-i\Sigma^<(x,p) &=& \sum_k \int \frac{d^4p_1}{(2\pi)^4} \frac{d^4p_2}{(2\pi)^4}
(2\pi)^4 \delta^4(p - p_1 - p_2)
\\ \nonumber & & \hspace{4cm} \times
i\Gamma_k \, iS^<(x,p_1) i\Gamma_k^\dagger i \Delta_k^>(x, -p_2) \Punkt
\eea
The gain term of Eq.~(\ref{qtrans}) thus reads
\bea
{\rm Tr} \left[ \Sigma^<(p) S^>(p) \right] &=& \sum_k
\int \frac{d^4p_1}{(2\pi)^4} \frac{d^4p_2}{(2\pi)^4}
(2\pi)^4 \delta^4(p - p_1 - p_2) i \Delta_k^>(-p_2)
\label{colli2} \\ \nonumber & & \hspace{4cm} \times
{\rm Tr} \left[ i\Gamma_k \, iS^<(p_1) \, i\Gamma_k^\dagger iS^>(p) \right]
\Punkt
\eea
In this expression, the off-diagonal meson Green function $\Delta_k^>$
appears. For the calculation of $\Sigma^<$, one has to take into account
both the singular and the regular part of this function. The two
contributions of $\Delta_{k,s}^>$ and $\Delta_{k,r}^>$ will be computed
separately in the following.

\subsubsection{Collision Term from the Singular Part}
For simplicity, only the contribution of the neutral pion is computed
explicitly. The other contributions can be evaluated along the same lines
and give similar results. To compute the contribution of the singular
part, one has to insert Eqs.~(\ref{KBansatz}) and (\ref{slmeson})
into Eq.~(\ref{colli2}). Since for $S^>(p)$ only the contribution
at positive $p_0$ is relevant, multiplying out Eq.~(\ref{colli2}) again
leads to four contributions, which can be identified with the processes
$q\to q\pi^0$, $q\pi^0\to q$, $\emptyset\to q\bar q\pi^0$ and $\pi^0\to
q\bar q$.  As was the case for the meson collision term, only the last
term survives for $m_\pi>2m_q$ due to kinematical constraints. The
contribution of this process reads, after a substitution $p_1\to -p_1$,
\bea
& & {\rm Tr} \left[ \Sigma^<(p) S^>(p) \right]_s =
\\ \nonumber & & \hspace{1cm}
\int \frac{d^4p_1}{(2\pi)^4} \frac{d^4p_2}{(2\pi)^4}
(2\pi)^4 \delta^4(p + p_1 - p_2)
\\ \nonumber & & \hspace{1cm} \times
\left[ g_\pi^2 \frac{\pi}{E_\pi(\vec p_2)} n_{\pi^0}(\vec p_2)
\delta(p_{20}-E_\pi(\vec p_2)) \right]
\\ \nonumber & & \hspace{1cm} \times
{\rm Tr} \Bigg\{ \gamma_5\tau_3
\left[ \frac{\pi}{2E_q(\vec p_1)} \frac{\delta_{ff'}\delta_{cc'}}{N_cN_f}
\left(-\sla p_1+m_q\right) \delta\left(p_{10}-E_q(\vec p_1)\right)
\left(2N_cN_f-n_{\bar q}(\vec p_1)\right)\right]
\\ \nonumber & & \hspace{1.5cm} \times
\gamma_5\tau_3
\left[ \frac{\pi}{2E_q(\vec p)} \frac{\delta_{ff'}\delta_{cc'}}{N_cN_f}
\left(\sla p+m_q\right) \delta\left(p_{0}-E_q(\vec p)\right)
\left(2N_cN_f-n_q(\vec p)\right)\right] \Bigg\}
\\ \nonumber
&=& \frac{\pi}{E_q(\vec p)} \delta \left(p_{0}-E_q(\vec p)\right)
\int \frac{d^3p_1}{(2\pi)^3 2E_q(\vec p_1)}
\frac{d^3p_2}{(2\pi)^3 2E_\pi(\vec p_2)}
(2\pi)^4 \delta^4(p+p_1-p_2)
\\ \nonumber & & \hspace{1cm} \times
\overline{\left| {\cal M} \right|^2}
n_{\pi^0}(\vec p_2) \left(2N_cN_f-n_q(\vec p)\right)
\left(2N_cN_f-n_{\bar q}(\vec p_1)\right) \Komma
\eea
where $\overline{\left| {\cal M} \right|^2}$ is again given by
Eq.~(\ref{tramp}). The total contribution of the singular part
$\Delta_{0,s}^{>,<}$ to the collision term is thus given by
\bea
& &
\int_0^\infty \frac{dp_0}{2\pi} {\rm Tr} \left[ \Sigma^<(p) S^>(p)
-\Sigma^>(p) S^<(p) \right]_s =
\\ \nonumber & & \hspace{8mm}
\frac{1}{2E_q(\vec p)}
\int \frac{d^3p_1}{(2\pi)^3 2E_q(\vec p_1)}
\frac{d^3p_2}{(2\pi)^3 2E_\pi(\vec p_2)}
(2\pi)^4 \delta^4(p+p_1-p_2) \overline{\left| {\cal M} \right|^2}
\\ \nonumber & & \hspace{16mm} \times
\left\{ n_{\pi^0}(\vec p_2) \left[2N_cN_f-n_q(\vec p)\right]
\left[2N_cN_f-n_{\bar q}(\vec p_1)\right] - n_q(\vec p) n_{\bar q}(\vec p_1)
\left[1+n_{\pi^0}(\vec p_2)\right] \right\}
\eea
and describes the process $\pi^0 \leftrightarrow q\bar q$.

\subsubsection{Collision Term from the Regular Part}
In order to obtain the contribution from the regular part of
$\Delta_k^{>,<}$ to the collision term, one has to insert
\be \label{vier12}
\Delta_{k,r}^{>,<}(x,p) = \frac{1}{2} \Pi_k^{>,<}(x,p) \left[
\Delta_k^R(x,p)^2 + \Delta_k^A(x,p)^2\right]
= \Pi_k^{>,<}(x,p) \Re \left(\Delta_k^R(x,p)^2\right)
\ee
into Eq.~(\ref{colli2}). The polarization $\Pi_k^{>,<}(x,p)$ here has
to be taken from Eq.~(\ref{vollpolar}). This leads, after performing
one of the integrations by means of the delta function, to
\bea
& & {\rm Tr} \left[ \Sigma^<(p) S^>(p) \right]_r
\\ \nonumber & & \hspace{2cm} = \int
\frac{d^4p_1}{(2\pi)^4} \frac{d^4p_2}{(2\pi)^4} \frac{d^4p_3}{(2\pi)^4}
(2\pi)^4 \delta^4(p-p_1+p_2-p_3) \Re \left(\Delta_k^R(p_2-p_3)^2\right)
\\ \nonumber & & \hspace{30mm} \times
{\rm Tr} \left[ i\Gamma_0 \, iS^<(p_1) \, i\Gamma_0 \, iS^>(p) \right]
{\rm Tr} \left[ i\Gamma_0 \, iS^>(p_2) \, i\Gamma_0 \, iS^<(p_3) \right]
\Punkt
\eea
The next part of the calculation proceeds exactly like shown in
Ref.~\cite{ogu}. Multiplying out the contributions of the individual quark
Green functions leads to the appearance of eight terms, of which five
can be dropped since they correspond to processes like $q\leftrightarrow
qq\bar q$ and $\emptyset\leftrightarrow q\bar qq\bar q$. The remaining
three terms can be rearranged to give
\bea
& & {\rm Tr}\left[ \Sigma^>(p) S^<(p) \right]_r = \frac{\pi}{E_q(\vec p)}
\delta(p_0-E_q(\vec p)) \label{stossterm}
\\ \nonumber & & \times
\int
\frac{d^3p_1}{(2\pi)^3 2E_q(\vec p_1)}
\frac{d^3p_2}{(2\pi)^3 2E_q(\vec p_2)}
\frac{d^3p_3}{(2\pi)^3 2E_q(\vec p_3)}
(2\pi)^4\delta^4(p+p_1-p_2-p_3)
\\ \nonumber & & \times
\bigg\{
\frac{1}{2} \left[ \phi(p_1-p_3) + \phi(p_1-p_2)\right]
n_q(p_2) n_q(p_3) \left[2N_cN_f-n_q(p_1)\right]
\left[2N_cN_f-n_q(p)\right]
\\ \nonumber & & \hspace{5mm}
+\left[\phi(p_1-p_3) + \phi(-p_2-p_3) \right]
n_q(p_2) n_{\bar q}(p_3) \left[2N_cN_f-n_{\bar q}(p_1)\right]
\left[2N_cN_f-n_q(p)\right] \bigg\} \Komma
\eea
where $\phi(p)$ is a shorthand notation for
\be \label{falsch}
\phi(p) = \frac{1}{(2N_cN_f)^2} p^4 \Re\left(\Delta_0^R(p)^2\right)
\Punkt
\ee
Equation (\ref{stossterm}) would have a form corresponding to a gain
term due to elastic quark-quark and quark-antiquark scattering, if one
could identify the collision kernel $\phi$ with the squared amplitude of
the diagrams shown in Fig.~\ref{scatfig}. In this case $\phi(p_1-p_3)$ would
correspond to the $t$-channel exchanges of Fig.~\ref{scatfig}a and d,
$\phi(p_1-p_2)$ to the $u$-channel exchange of Fig.~\ref{scatfig}b and
$\phi(-p_2-p_3)$ to the $s$-channel exchange of Fig.~\ref{scatfig}c. Note
that the interference terms of these diagrams belong to a higher order
in $1/N_c$ \cite{elaste} and do not arise from the self-energy shown in
Fig.~\ref{sigmafig}b, as has been detailled in Ref.~\cite{ogu}.

The problem with this interpretation is, that a direct calculation of
the squared transition amplitudes corresponding to the graphs shown in
Fig.~\ref{scatfig} shows, that the collision kernel obtained from these
is not given by Eq.~(\ref{falsch}), but rather by \cite{elaste}
\be
\label{richtig} \Phi(p) =
\frac{1}{(2N_cN_f)^2} p^4 \left|\Delta_0^R(p)\right|^2 \Punkt
\ee
For the $t$ and $u$-channel exchanges, this makes no difference,
since in these cases $\Delta_0^R(p)$ is real \cite{elaste} and thus
Eqs.~(\ref{falsch}) and (\ref{richtig}) yield the same result. A problem
arises, however, with the $s$-channel, for which Eq.~(\ref{falsch}) gives
effectively the wrong exchange propagator.  The results of this effect
can be seen from Fig.~\ref{effpfig}. The solid line in this figure shows
the square of the full exchange propagator, $\left|\Delta_0^R\right|^2$,
for $s$-channel exchange as a function of $\sqrt{s}$ at thermal
equilibrium and $T=300$~MeV, whereas
the dashed line shows its replacement in Eq.~(\ref{falsch}),
$\Re({\Delta_0^R}^2)$. Since the latter leads to a negative collision
kernel at large $\sqrt{s}$, Eq.~(\ref{falsch}) cannot be regarded as a
valid collision kernel for $s$-channel scattering.

The physical reason for this breakdown of the theory can be seen from
the following: the full transition amplitude (\ref{richtig}) contains
resonance exchange, i.\,e. contributions of the process $q\bar q\to\pi\to
q\bar q$. In the present approach, however, these contributions have to be
taken out from the elastic scattering amplitude, since they are already
counted in the collision terms for $q\bar q\leftrightarrow \pi$. Using
the full transition amplitude (\ref{richtig}) is thus double counting.

It is thus to be expected, that the $s$-channel elastic scattering
amplitude has to be modified near a resonance. Nevertheless,
Eq.~(\ref{falsch}) cannot be the correct form since it (i) gives
a negative collision kernel, (ii) modifies the collision kernel at
momenta far off from the resonance, where no modification is to be
expected and (iii) gives also the wrong results at low temperatures,
where the production of pions via recombination is kinematically
forbidden and thus no interference can take place.

The form (\ref{richtig}) for the collision kernel can be obtained by
setting \cite{davis,bedu}
\be \label{beduine}
\Delta_k^{>,<} \approx \Pi_k^{>,<} \left| \Delta_k^R \right|^2 \Punkt
\ee
This ansatz, however, inhibits the appearance of mesons as individual
degrees of freedom. A similar result was obtained in Ref.~\cite{davis},
where it was shown that a quasiparticle ansatz for the meson
propagator leads to collision terms containing processes of the type
$q\bar q\leftrightarrow \pi$, but no elastic scattering of quarks,
whereas the ansatz (\ref{beduine}) leads to the elastic scattering
amplitude (\ref{richtig}), eliminates however the appearance of free
mesons. Note that this problem is not a generic problem of a model
containing mesons as bound states, but rather a general problem of
theories containing interacting quarks and mesons, as can be concluded
from Ref.~\cite{davis}. Nevertheless, elastic scattering processes {\em
have\/} to be obtained from the diagram \ref{sigmafig}b. They cannot be
derived from self energy graphs like the one shown in Fig.~\ref{unsinn},
which would lead to the appearance of elastic scattering processes
when evaluated in a quasiparticle approximation: This graph is already
implicitly included in the graph shown in Fig.~\ref{sigmafig}b via the
meson line and its explicit appearance would thus be double counting.

The conclusion from this is, that considering off-shell corrections
to $\Delta_k^{>,<}$ is essential in order to obtain collision terms
containing elastic scattering at all, the form of the off-shell
corrections given in Ref.~\cite{pavel} is, however, insufficient as
far as the $s$-channel scattering amplitude is concerned.

In order to find a more suitable form of the off-shell corrections,
one has to consider a different decomposition of $\Delta_k^{>,<}$ into
a singular and regular part as that of Eqs.~(\ref{drei13}). An attempt
to do this, which however leads to the appearance of new terms in the
derivation of the Boltzmann equation, is shown in the appendix. This
decomposition leads to the alternative form
\be
\Delta_{k,r}^{>,<} =\Pi_k^{>,<}
\left[ \left|\Delta_k^R\right|^2 - 2\left(\Im\Delta_{k,p}^R \right)^2\right]
\label{interpol}
\ee
for the regular part of $\Delta_k^{>,<}$. The auxiliary quantity
$\Delta_{k,p}^R$ in Eq.~(\ref{interpol}) is defined by
\be
\Delta_{k,p}^R(x,p) = \frac{g_k^2(x, \vec p)}{2 E_k(x,\vec p)
\left( p_0 - E_k(x,\vec p) + \frac{i}{2} \Gamma_k(x,\vec p) \right)}
\Punkt
\ee
This form of the off-shell corrections improves (\ref{vier12}) in so far,
as the effective exchange propagator is modified only in a small region
around the meson mass. If one has $\Delta_{k,p}^R\approx\Delta_k^R$, one
reobtains the form (\ref{falsch}) for the collision kernel. Far off from
the pole, $\Delta_{k,p}^R$ can be neglected and one obtains the collision
kernel (\ref{richtig}), which in this region is dominated by scattering
states. Equation (\ref{interpol}) has thus the property to interpolate
between these two forms. It is nevertheless not a complete solution to
the problem, as can be seen from the dotted line in Fig.~\ref{effpfig},
which shows that Eq.~(\ref{interpol}) gives a good approximation to
Eq.~(\ref{richtig}) far off the pion mass. This is especially important in
order to obtain the correct scattering amplitude at low temperatures,
where the formation of pions by recombination is kinematically
forbidden. Near the meson mass the transition amplitude is modified,
as expected. Equation (\ref{interpol}) gives, however, still a negative
contribution at this energy and can thus not be regarded as a complete
solution to the problem. It is nevertheless worthwhile to look for
improved off-shell corrections by considering alternative decompositions.

\section{Summary and Conclusions} \label{conclusec}
In the preceding sections, it is shown that it is possible to extend the
formalism of transport theory in order to describe the evolution of a
system of quarks and bound state mesons. Each of these components
evolves according to a Boltzmann equation containing a mean field
term and collision integrals, which describe the interactions with the
other component. Since mesons appear as effective degrees of freedom,
the dynamical transition from a quark phase to a hadronical phase can
be modeled. The derivation of the transport equations leads to medium
dependent transition amplitudes in the collision integrals.

In order to obtain collision integrals for higher order processes,
the consideration of off-shell corrections to the off-diagonal Green
functions is essential. These off-shell corrections have to be chosen
carefully, since they modify the collision kernels in a nontrivial way.

In order to extend the current approach to a nonequilibrium description of
the chiral phase transition, further work has still do be done. Although
the present approximation allows for the transformation of quark pairs
into hadrons via the process $q\bar q\to\pi$, this process cannot be
expected to be very efficient, since its probability is limited by the
phase space. One has thus to consider the processes $q\bar q\to\pi\pi$
\cite{hadron} or $q\bar q\to\pi\pi\pi$ \cite{dagmar}, which can proceed
via a much less limited kinematics. This in turn demands the inclusion
of higher order self energies \cite{hirsch}. For this work, it might
also be necessary to introduce off-shell corrections for the quark
Green functions.

One of the requirements for this expansion is that it has to be symmetry
conserving. This concerns especially chiral symmetry and the validity of
the Goldstone theorem. A symmetry conserving expansion for the equilibrium
case has been outlined in Ref.~\cite{richard}. A generalization of this
work to non-equilibrium and the derivation of collision terms for this
expansion is left for a future publication.

\section*{Acknowledgments}
Discussions with S.\,P.~Klevansky, J.~H\"ufner, J.~Aichelin and
P.~Lipavsk\'y are gratefully acknowledged. This work has been supported in
part by the Deutsche Forschungsgemeinschaft under contract no.~Hu~233/4-4,
and the German Federal Ministry for Education and Research under contract
no. 06~HD~742.

\begin{appendix}
\section*{Derivation of Improved Off-Shell Corrections}
The pole approximation of the retarded Green function given in
Eq.~(\ref{pola}) reads
\be
\Delta_k^R(x,p) \approx \label{a1}
\frac{g_k^2(x, \vec p) - i a_k(x, \vec p)}{2 E_k(x,\vec p)
\left( p_0 - E_k(x,\vec p) + \frac{i}{2} \Gamma_k(x,\vec p) \right)}
\Punkt
\ee
The imaginary part $a_k(x, \vec p)$ of the pole residue appearing
here is of the order of the width and thus gives no contribution
to the final result in Section~\ref{mdersec}. It has also not been
considered in Ref.~\cite{pavel}. In the following it will thus be attempted
to derive transport equations using the effective form
\be
\Delta_{k,p}^R(x,p) = \label{a2}
\frac{g_k^2(x, \vec p)}{2 E_k(x,\vec p)
\left( p_0 - E_k(x,\vec p) + \frac{i}{2} \Gamma_k(x,\vec p) \right)}
\ee
for the retarded Green functions. To do this, define the auxiliary
quantities
\bem \label{a3} \bea
\rho_{k,p}(x,p) &=& i\left(\Delta_{k,p}^R(x,p) - \Delta_{k,p}^A(x,p) \right)
\\
\Delta_b^{-1} - \Pi_{k,p}^R(x,p) &=& \frac{2E_k(x, \vec p)}{g_k^2(x, \vec p)}
\left(p_0-E_k(x, \vec p) + \frac{i}{2}\Gamma_k(x,\vec p) \right)
\\
\Pi_{k,p}(x,p) &=& \frac{1}{2} \left( \Pi_{k,p}^R(x,p) +
\Pi_{k,p}^A(x,p)\right)
\\
\gamma_{k,p}(x,p) &=& i \left( \Pi_{k,p}^R(x,p) - \Pi_{k,p}^A(x,p)\right)
\Punkt
\eea \eem
It can be easily verified, that after a transformation to coordinate
space $\Delta_{k,p}^{R,A}$ fulfill the equations
\bem \bea
\left( \Delta_b^{-1} \otimes \Delta_{k,p}^{R,A} \right) (x, y) &=&
\delta^4(x-y)
+ \left( \Pi_{k,p}^{R,A} \otimes \Delta_{k,p}^{R,A} \right) (x,y)
\\
\left( \Delta_{k,p}^{R,A} \otimes \Delta_b^{-1} \right) (x, y) &=&
\delta^4(x-y)
+ \left( \Delta_{k,p}^{R,A} \otimes \Pi_{k,p}^{R,A} \right) (x,y)
\eea \eem
and $\rho_{k,p}$ obeys the equation
\be
\rho_{k,p} = \Delta_{k,p}^R \otimes \gamma_{k,p} \otimes \Delta_{k,p}^A
\Punkt
\ee
The decomposition of $\Delta_k^{>,<}$ and $\rho_k$ is done such that
the singular parts are defined as
\bem \bea
\Delta_{k,s}^{>,<} &=&  \frac{i}{2} \left(\Delta_{k,p}^R \otimes \Pi_k^{>,<}
                        \otimes \rho_{k,p} - \rho_{k,p} \otimes \Pi_k^{>,<}
                        \otimes \Delta_{k,p}^A \right)
\\
\rho_{k,s} &=& \frac{i}{2} \left(\Delta_{k,p}^R \otimes \gamma_k \otimes
               \rho_{k,p} - \rho_{k,p} \otimes \gamma_k \otimes
               \Delta_{k,p}^A \right)
\eea \eem
and the regular parts are defined as the difference of the full function
and the singular part.

Analogously to Eq.~(\ref{way1}), one obtains
\bea
& &-i \left[ \left(\Delta_{k,p}^R\right)^{-1} \otimes \Delta_{k,s}^<
- \Delta_{k,s}^< \otimes  \left(\Delta_{k,p}^A\right)^{-1} \right]
= \frac{1}{2} \left( \gamma_{k,p} \otimes \Delta_{k,s}^<
+ \Delta_{k,s}^< \otimes \gamma_{k,p} \right)
\\ \nonumber & & \hspace{3cm}
 - i \left[ \left( \Delta_b^{-1} - \Pi_{k,p} \right) \otimes \Delta_{k,s}^<
 - \Delta_{k,s}^< \otimes \left( \Delta_b^{-1} - \Pi_{k,p} \right) \right]
\eea
and Eq.~(\ref{way2}) is replaced by
\bea
& &-i \left[ \left(\Delta_{k,p}^R\right)^{-1} \otimes \Delta_{k,s}^<
- \Delta_{k,s}^< \otimes  \left(\Delta_{k,p}^A\right)^{-1} \right]
= \frac{1}{2} \left( \Pi_k^< \otimes \rho_{k,p} + \rho_{k,p}
\otimes \Pi_k^< \right)
\\ \nonumber & & \hspace{3cm}
- \frac{1}{2} \left(
\gamma_{k,p} \otimes \Delta_{k,p}^A \otimes \Pi_k^< \otimes \Delta_{k,p}^A
+ \Delta_{k,p}^R \otimes \Pi_k^< \otimes \Delta_{k,p}^R \otimes \gamma_{k,p}
\right) \Punkt
\eea
After performing the Wigner transformation, this becomes
\bea
& & \left\{ \Delta_b^{-1}-\Pi_{k,p} ; \Delta_{k,s}^< \right\} +
\frac{1}{4} \left\{ \Pi_k^< \left( \Delta_{k,p}^R + \Delta_{k,p}^A \right)
\rho_{k,p} ; \gamma_{k,p} \right\}
\label{a9} \\ \nonumber & & \hspace{4cm} = \Pi_k^<
\left[\rho_{k,p} - \frac{1}{2} \gamma_p \left( {\Delta_{k,p}^R}^2 +
{\Delta_{k,p}^A}^2 \right) \right] - \gamma_{k,p} \Delta_{k,s}^<
\Punkt
\eea
By using the explicit representation (\ref{a2}) and the definitions
(\ref{a3}), it is possible to show that
\be
\rho_{k,p} - \frac{1}{2} \gamma_p \left( {\Delta_{k,p}^R}^2 +
{\Delta_{k,p}^A}^2 \right) = \tilde \rho_{k,s} = \frac{1}{2}
\gamma_{k,p} \rho_{k,p}^2 \Punkt
\ee
After employing the factorization $\Delta_{k,s}^<=-i\rho_{k,s} f_k$,
the first term on the left hand side of Eq.~(\ref{a9}) becomes
\bea
\left\{ \Delta_b^{-1}-\Pi_{k,p} ; \Delta_{k,s}^< \right\}
&=& -i \rho_{k,s} \left\{ \Delta_b^{-1}-\Pi_{k,p} ; f_k \right\}
- i f_k \left\{ \Delta_b^{-1}-\Pi_{k,p} ; \rho_{k,s} \right\}
\\ \nonumber
&=& -i \rho_{k,s} \left\{ \Delta_b^{-1}-\Pi_{k,p} ; f_k \right\}
- i f_k \left\{ \Delta_b^{-1}-\Pi_{k,p} ;
\frac{1}{2}\left(\gamma_k-\gamma_{k,p} \right) \rho_{k,p}^2 \right\}
\\ \nonumber & &
- i f_k \left\{ \gamma_{k,p} ; \frac{1}{4}\gamma_{k,p}
\left( \Delta_{k,p}^R + \Delta_{k,p}^A \right) \rho_{k,p} \right\}
\Komma
\eea
which generalizes Eq.~(\ref{drei23}) to the present case. Repeating the
steps which led to the derivation of Eq.~(\ref{mesoboltz}) leads now to
\bea \label{a12}
& & - i \rho_{k,s} \left\{ \Delta_b^{-1}-\Pi_{k,p} ; f_k \right\}
- i f_k \left\{ \Delta_b^{-1}-\Pi_{k,p} ;
\frac{1}{2}\left(\gamma_k-\gamma_{k,p} \right) \rho_{k,p}^2 \right\}
\\ \nonumber & & \hspace{10mm}
+ i \tilde\rho_{k,s} \frac{\Delta_b^{-1} - \Pi_{k,p}}{\gamma_{k,p}}
\left\{\gamma_{k,p} ; f_k \right\}
- \left\{ \gamma_{k,p} ; \frac{1}{4} \rho_{k,p}
\left( \Delta_{k,p}^R + \Delta_{k,p}^A \right) \left(if_k\gamma_{k,p}
+\Pi_k^<\right) \right\}
\\ \nonumber & & \hspace{20mm}
= \Pi_k^< \tilde \rho_{k,s} + i \gamma_{k,p} f_k \rho_{k,s} \Punkt
\eea
The fourth term on the left hand side can be dropped using the same
arguments as before. The second term on the left hand side is new.
Since it is proportial to the gradient of the difference
$\gamma_k-\gamma_{k,p}$, it is also regarded as belonging to higher
orders and thus neglected. The remaining terms contain either a factor
$\rho_{k,s}$ or $\tilde\rho_{ks}$. Although these are not equal, they
are of the same order and become strongly peaked near
the quasiparticle energy in the small width limit. It is thus justified
to drop these factors and one obtains
\be
\left\{ \Delta_b^{-1}-\Pi_{k,p} ; f_k \right\}
- \frac{\Delta_b^{-1} - \Pi_{k,p}}{\gamma_{k,p}}
\left\{\gamma_{k,p} ; f_k \right\}
= i \Pi_k^< - \gamma_{k,p} f_k \Komma
\ee
where all momenta have to be taken on shell. The second term on the left
hand side vanishes. Furthermore it can be inferred from Eqs.~(\ref{a2})
and (\ref{a3}), that up to the order $a_k^2$ one has $\gamma_{k,p} =
\gamma_k$ on the mass shell.  After identifying $f_k$ with the particle
density, one reobtains the Boltzmann Equation (\ref{drei30}).

Although this derivation gives the same form of the transport equation,
it differs in the form of the regular part of $\Delta_{k,r}^<$. By
definition, one has in coordinate space
\bea
\Delta_{k,r}^{>,<} = \Delta_k^{>,<} - \Delta_{k,s}^{>,<} &=&
\Delta_k^R \otimes \Pi_k^{>,<} \otimes \Delta_k^A
\\ \nonumber & &
-\frac{i}{2} \left( \Delta_{k,p}^R \otimes \Pi_k^{>,<} \otimes \rho_{k,p}
- \rho_{k,p} \otimes \Pi_k^{>,<} \otimes \Delta_{k,p}^A \right) \Punkt
\eea
Transforming this to momentum space and neglecting the gradient terms,
which are of the same order as the term neglected in Eq.~(\ref{a12}),
leads to
\be
\Delta_{k,r}^{>,<} =
\Pi_k^{>,<} \left( \Delta_k^R \Delta_k^A - \frac{1}{2} \rho_{k,p}^2 \right)
\Komma
\ee
which is equal to Eq.~(\ref{interpol}).
\end{appendix}

\begin{figure}
\caption[]{The Schwinger--Keldysh contour $C$ in the complex time plane.}
\label{contour}
\end{figure}

\begin{figure}
\caption[]{Pion spectral function as a function of $p_0$ at thermal
           equilibrium and $T=0$. The model parameters used for
           this calculation are $m_0=5$~MeV for the current quark
           mass, $G\Lambda^2=2.105$ for the coupling constant and
           $\Lambda=653$~MeV for the NJL cutoff, which gives a quark
           mass of $m_q=313$~MeV from Eq.~(\ref{gap}) and a pion mass
           of $m_\pi=134$~MeV from Eq.~(\ref{dispers}). Note the delta
           peaks at $p_0=\pm m_\pi$, which correspond to bound state
           pions. The continua for $|p_0|>2m_q$ stem from $q\bar q$
           scattering states.}
\label{spekt1}
\end{figure}

\begin{figure}
\caption[]{Pion spectral function as a function of $p_0$ at thermal
	   equilibrium and $T=300$~MeV. The model parameters are the
	   same as for Fig.~\ref{spekt1}, leading at this temperature to
	   $m_q=21$~MeV and $m_\pi=366$~MeV.  The solid line corresponds
	   to the full spectral function, the dashed line gives a pole
	   approximation. The delta peaks visible in Fig.~\ref{spekt1}
	   have now obtained a finite width and the pion has become a
	   resonant state.}
\label{spekt2}
\end{figure}

\begin{figure}
\caption[]{Lowest order diagram for the polarization $\Pi_k$. Solid
           lines correspond to quarks.}
\label{polafig}
\end{figure}

\begin{figure}
\caption[]{Lowest order diagrams for the quark self energy $\Sigma$.
           Solid lines correspond to quarks, double lines to mesons.
           Diagram (a) gives the Hartree self energy, diagram (b) the
           lowest order contribution to the collision term.}
\label{sigmafig}
\end{figure}

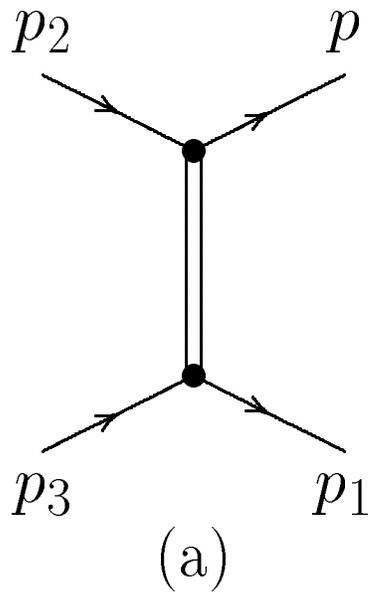
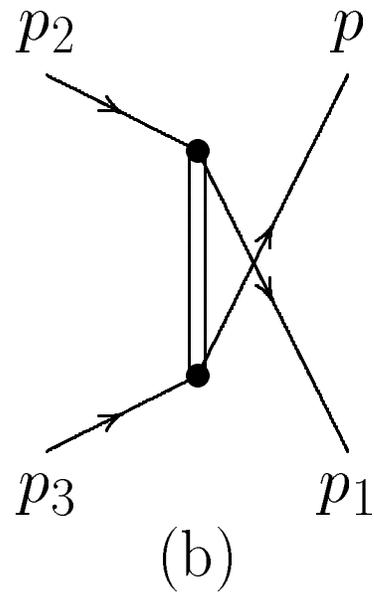
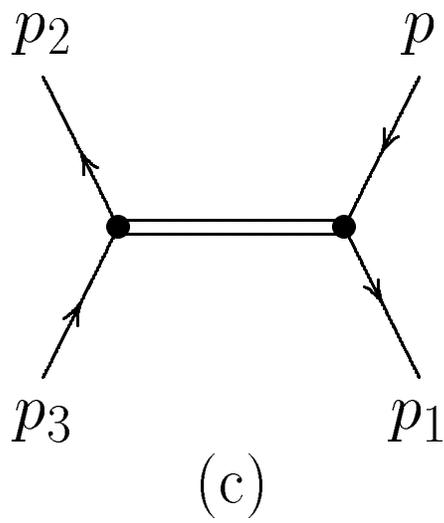
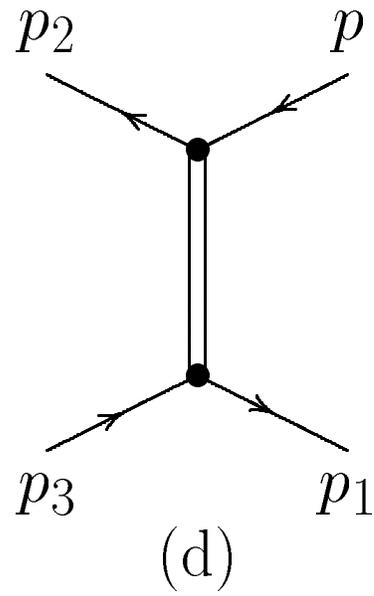
\begin{figure}
\caption[]{Lowest order diagrams for elastic scattering by meson
           exchange. Solid lines correspond to quarks, double lines
           to mesons. The diagrams (a) and (b) give the $t$ and $u$
           channel exchange for quark-quark scattering, (c) and (d) the
           $s$ and $t$ channel exchange for quark-antiquark scattering.}
\label{scatfig}
\end{figure}

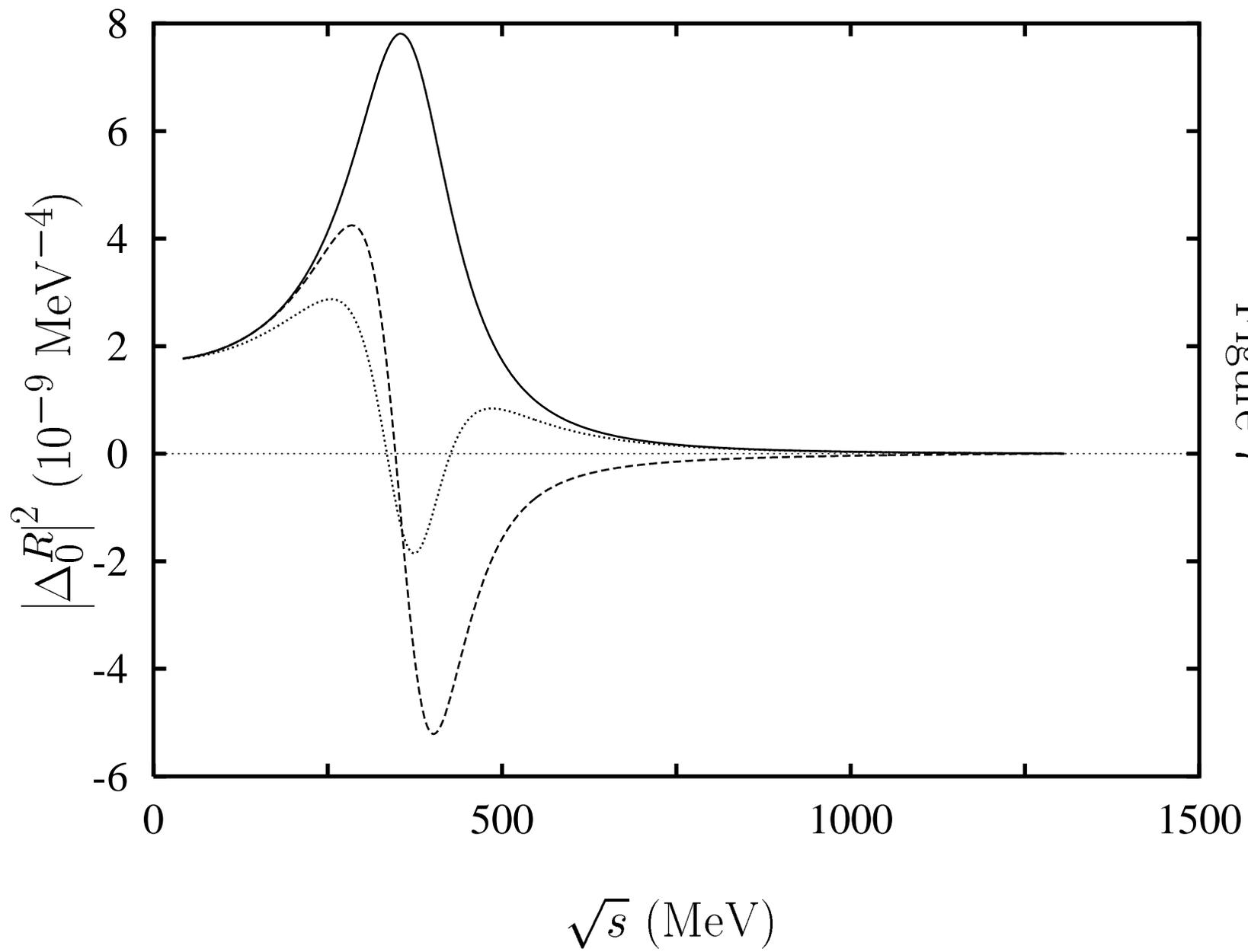
\begin{figure}
\caption[]{Exchange propagators for the $s$-channel of elastic
           quark-antiquark scattering in equilibrium at
           $T=300$~MeV as a function of $\sqrt{s}$. Solid
           line: $\left|\Delta_0^R\right|^2$, dashed line:
           $\Re\left({\Delta_0^R}^2\right)$, dotted line: interpolating
           form according to Eq.~(\ref{interpol}). The vertical axis is
           given in arbitrary units.}
\label{effpfig}
\end{figure}

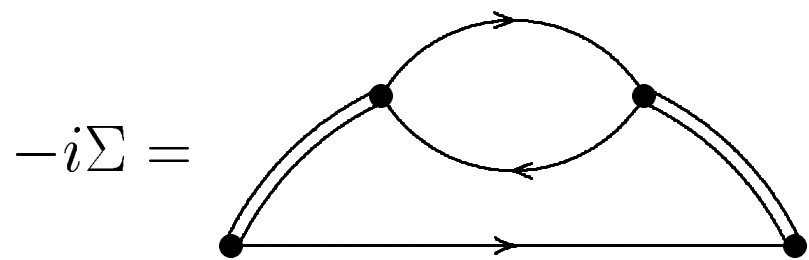
\begin{figure}
\caption[]{Self energy graph leading to elastic scattering processes when
           evaluated in a quasiparticle approximation. Solid lines
           correspond to quarks, double lines to mesons.}
\label{unsinn}
\end{figure}

\end{document}